\documentstyle[aps,epsfig]{revtex}
\begin{document}
\title{Quantum imaging} 
\author{L.A. Lugiato, A. Gatti and  E. Brambilla } 
\address{ Istituto Nazionale per la Fisica della Materia, Dipartimento di 
Scienze Chimiche Fisiche e Matematiche, Universit\`{a} dell'Insubria, Via Valleggio 11, 
22100 Como, Italy}
\maketitle

\begin{abstract}
We provide a brief overview of the newly born field of quantum imaging, and discuss some 
concepts that lie at the root of this field.
\end{abstract} 
\section{Preamble}
\label{Preamble}

This article is based on the plenary talk that one of us (L.A.L.) gave at the ICSSUR Conference in 
Boston in June 2001. \\
The starting point is provided by the general topic of the spatial aspects of quantum optical 
fluctuations. This has been the object of several studies in the past, but only recently there has been 
a constant focus of attention, which is mainly due to the fact that the spatial features may open new 
possibilities, e.g. in the direction of parallel processing and multichannel operation by quantum 
optical procedures. Once realized that these studies may have interesting perspectives, it is natural 
to coin a new name as quantum imaging to designate this field.\\
This article includes an introductory part, in which we discuss few concepts that play a key role in 
this field, such as the intrinsic connection between quantum entanglement and squeezing, the spatial 
squeezing and the near field/far field duality. Next, we discuss several topics in the field of quantum 
imaging and illustrate recent ideas and results. The topics include
\begin{description}
\item{-} Detection of weak phase and amplitude objects beyond the standard quantum limit,
\item{-} Amplification of weak optical images preserving the signal-to-noise ratio (noiseless amplification),
\item{-} Entangled two-photon microscopy,
\item{-} Quantum limits in the detection of small displacements and in image reconstruction,
\item{-} Quantum lithography,
\item{-} Quantum teleportation of optical images.
\end{description}
Several of the results and approaches that we will illustrate have been pursued by the participants in 
the European Project QUANTIM (Quantum Imaging), that started in January 2001.

\section{Introduction}
\label{Introduction}

\subsection{Intrinsic connection between squeezing and quantum entanglement}
\label{connection}
One might consider that squeezing is a rather old-fashioned field, 
whereas for entanglement one immediately thinks of such topics as 
quantum  computing, or quantum teleportation, or 
cryptography, and concludes that it is a new and exciting field. 
However one must remark that, in the framework of the continuous variable 
approach, squeezing and quantum entanglement are 
intrinsically linked, they are basically two faces of the same phenomenon. 
This circumstance explains by the way why a series of meetings 
entitled Squeezed States and Uncertainty Relations is 
still gathering so many people. \\
In order to demonstrate this point in details, let us consider 
two radiation beams and the associated 
annihilation operators of photons $a_1$ and $a_2$, and let us focus on 
the simple linear transformation to another couple of beams $b_1$ and $b_2$:
\begin{equation}
\label{atob}
b_1= \frac {a_1 +a_2} {\sqrt{2}}\:, \qquad b_2= \frac {a_1 - a_2} {\sqrt{2}}
\end{equation}
This transformation can be implemented very easily. If, for example, 
$a_1$ and $a_2$  have the same 
frequency and the same polarization, it is realized by a 50/50 beam splitter. 
If they have the same frequency  but horizontal and vertical polarization, 
respectively, one can use a polarizing beam 
splitter in which the two output beams are polarized at $45^0$ and  $135^0$, 
respectively.\\ 
Now, the general result for transformation (1) is that
\begin{description}
\item{--} if $a_1$ and $a_2$  are  EPR 
(Einstein-Podolski-Rosen \cite{[1]}) entangled beams 
with respect to quadrature components, 
then beams  $b_1$ and $b_2$ are squeezed with respect to two orthogonal 
quadrature components and, vice versa,
\item{--} if $a_1$ and $a_2$  are squeezed beams with respect to 
two orthogonal quadrature components,
$b_1$ and $b_2$ are EPR entangled beams with respect to 
quadrature components.
\end{description}
To prove this result, let us consider the interaction Hamiltonian 
for parametric down-conversion 
in the nondegenerate configuration and in the approximation of 
classical undepleted pump \cite{[2]}
\begin{equation}
H= i g \alpha \left(a_1^\dagger a_2^\dagger - a_1 a_2\right)
				\label{NDH}
\end{equation}
where $g$ is the coupling constant and $\alpha$ is the classical 
amplitude of the pump beam. As it is 
well known, in the nondegenerate configuration the two beams    
$a_1$ and $a_2$  are entangled both 
with respect to photon number and with respect to two 
quadrature components \cite{[2],[3],[4]}.
The entanglement can be explicitly shown by applying the time evolution operator corresponding
to the Hamiltonian (\ref{NDH}) to the uncorrelated vacuum state, for an 
interaction time $\tau_{int}$
\begin{eqnarray}
{\rm e}^{-\frac{i}{\hbar}H \tau_{int}} \: |0\rangle_1 |0\rangle_2   &=&  
\sum_{n=0}^{\infty} c_n \, |n \rangle_1 |n \rangle_2  \nonumber \\
c_n &=& \frac{\left[\tanh  \left( g \tau_{int} / \hbar\right)\right]^n }{ \cosh \left( g \tau_{int} / \hbar\right) }  \; . 
				\label{two-mode}
\end{eqnarray}
Clearly the state described by Eq.(\ref{two-mode}) is not factorizable, and implies perfect 
correlation between the photon number in the two beams.

If we now 
introduce into Eq.~(\ref{NDH}) the expression of $a_1$ and $a_2$  
as a function of  $b_1$ and $b_2$   
\begin{equation}
\label{btoa}
a_1= \frac {b_1 +b_2} {\sqrt{2}}\:, \qquad a_2= \frac {b_1 - b_2} {\sqrt{2}} \:,
\end{equation}
we obtain immediately the alternative expression
\begin{equation}
H= i \frac{g \alpha}{2} \left[ \left( b_1^\dagger \right)^2 - b_1^2\right]
     -i \frac{g \alpha}{2} \left[ \left( b_2^\dagger \right)^2 - b_2^2\right]\:,
				\label{DH}
\end{equation}
which corresponds to the sum of two interaction hamiltonians 
for parametric down-conversion in 
the degenerate configuration. 
Hence the beams $b_1$ and $b_2$ are squeezed, and, because of the       
minus sign in front of the second term,  
the squeezing is in orthogonal quadrature components. For 
example, when $\alpha$ is real $b_1$ is squeezed with respect 
to the Y quadrature (imaginary part of 
the annihilation operator) and $b_2$ is squeezed 
with respect to the X quadrature (real part of the 
annihilation operator).

\subsection{Spatially multimode versus singlemode squeezing}
\label{multisqueeze}

In almost all literature on squeezing one considers singlemode squeezing. If one wants to detect a 
good level of squeezing, the local oscillator must be matched to the squeezed spatial mode and, in 
addition, it is necessary to detect the whole beam. If one detects only part of the beam the squeezing 
is immediately degraded, because a portion of a mode  necessarily involves higher order modes, in 
which squeezing is absent. What we can call {\em local squeezing}, i.e. squeezing in small regions of the 
transverse plane, can be obtained only in presence of {\em spatially multimode squeezing}, i.e. squeezing 
in a band of spatial modes. This has been predicted by Sokolov and Kolobov for the traveling wave 
optical parametric amplifier (OPA) \cite{[5],[6]}  and by our group for the optical parametric oscillator 
(OPO) \cite{[7],[8]}.\\
Let us dwell a moment, for example, on the case of the OPA (Fig.1a), in which one has a slab of $\chi^{(2)}$             
material which is pumped by a coherent plane wave of frequency   $2\omega_s$. A fraction of the pump 
photons are down-converted into signal-idler photon pairs, which are distributed over a broad band 
of temporal frequencies around the degenerate frequency $\omega_s$. For each fixed temporal frequency, 
the photon pairs are distributed over a band of spatial frequencies labeled by the transverse 
component  $\vec{q}$ of the wave vector.\\
If, in addition to the pump field, we inject a coherent plane wave with frequency  $\omega_s$ and 
transverse wave vector $\vec{q}$ (Fig. 1b), in the output we have a signal wave which corresponds to an 
amplified version of the input wave and for this reason the system is called optical parametric 
amplifier. Because of the pairwise emission of photons, there is also an idler wave which, close to 
degeneracy, is symmetrical with respect to the signal wave.\\
Referring to the case in which only the pump is injected, two regimes can be distinguished. One is 
that of pure spontaneous parametric down-conversion, as in the case of a very thin crystal. 
In this case coincidences between partners of single photon pairs are detected. The other is that of 
dominant stimulated parametric down-conversion, in which a large number of photon 
pairs at a time is detected. In the following we will consider both cases alternatively.
\subsection{Near field/Far field duality}
\label{nearfar}
We want to illustrate the key spatial quantum properties of the field emitted by an OPA, in the 
linear regime of negligible pump depletion, or by an OPO below threshold. In the OPA case, we 
consider the configuration of a large photon number.\\
In the near field (see Fig. 2) one has the phenomenon of spatially multimode squeezing or local 
squeezing discussed in Sec.~\ref{multisqueeze}. A good level of squeezing is found, 
provided the region which is detected 
has a linear size not smaller than the inverse of the spatial bandwidth of emission in the 
Fourier plane. If, on the other hand, one looks at the far field (which can be reached, typically, by 
using a lens as shown in Fig.2) one finds the phenomenon of {\em  spatial entanglement}
between small regions located symmetrically with respect to the center. 
Precisely, if one considers two symmetrical 
pixels 1 and 2 (Fig.3), the intensity fluctuations in the two pixels  are very well  correlated or, 
equivalently, the fluctuations in the intensity difference between the two pixels are very much 
below the shot noise \cite{[9],[10]}. Because this phenomenon arises for any pair of symmetrical pixels, 
we call it spatial entanglement. The same effect occurs also for quadrature components, because 
in the two pixels the fluctuations of the quadrature component $X$ are almost exactly correlated, and 
those of the quadrature component $Y$  are almost exactly anticorrelated \cite{[11]}. The minimum size of 
the symmetrical small regions, among which one finds spatial entanglement, is determined by the 
final aperture of optical elements,
and is given, in the paraxial approximation, by   $\lambda f/a$, where $\lambda$ is the wavelength,
$f$ is the focal length of the lens and  $a$   is the aperture of optical elements (e.g. the lens aperture)(Fig. 2).
In a more realistic model of the OPA, the finite waist of the pump field should be taken into account.
In this case the minimum size of the regions where entanglement is detectable in the far field
is mostly determined by the pump waist.\\
The spatial entanglement of intensity fluctuations in the far field is quite evident even in  single 
shots (the pump field is typically pulsed).  Fig. 4a  shows a numerical simulation  in a case of non 
collinear phase matching at degenerated  frequency. One observes the presence of symmetrical 
intensity peaks, which become  broader and broader as one reduces the waist of the pump field. 
A similar situation is observed in an experiment performed  using a LBO crystal \cite{[12]}.
We observe finally that the near field/far field duality can be understood on the basis of the intrinsic 
connection between squeezing and quantum entanglement. The spatial entanglement in the far field 
arises from the correlation between the  modes $a_{\vec{q}} \sim \exp{ [i \vec{q} \cdot \vec{x}]}$  and
$a_{-\vec{q}} \sim \exp{ [-i \vec{q} \cdot \vec{x}]}$,                     , 
which in the far field give rise to two separated and opposite spots in the transverse plane  ($\vec{q}$ is 
the transverse wave vector and $\vec{x}$ is the position vector in the transverse plane).\\
On the other hand, in the near field there is no squeezing in modes   $a_{\vec{q}}$ and $a_{-\vec{q}}$                
separately, whereas there is large squeezing in the combination modes                                                
and   $b_{\vec{q}} = \left( a_{\vec{q}} + a_{-\vec{q}}\right) /\sqrt{2}$ and   $b_{-\vec{q}} = \left( a_{\vec{q}} - a_{-\vec{q}}\right) /\sqrt{2}$ ,
which annihilate photons on spatial modes $\sim \cos (\vec{q}\cdot\vec{x}) $ and $\sim \sin (\vec{q}\cdot\vec{x})$.
In the near field it is possible to observe this squeezing by 
using  a local oscillator with a  $ \cos (\vec{q}\cdot\vec{x}) $ or $ \sin (\vec{q}\cdot\vec{x})$
spatial configuration \cite{[7]}. One notices 
immediately that the relation between modes  $a_{\vec{q}},\: a_{-\vec{q}}$ and modes   $b_{\vec{q}},\: b_{-\vec{q}}$                    
coincides with Eq. (1), which, as we have seen, transforms  entangled beams into squeezed beams, 
and viceversa.           
\section{Topics in quantum imaging}
\label{Topics}
Next we focus specifically on studying the quantum aspects of the very classical field of imaging.\\ 
We will discuss several topics in order. 
\subsection{Detection of weak amplitude or phase objects beyond the standard quantum limit}
\label{weak }

Let us consider first the case of a weak amplitude object which is located, say, in the signal part of 
the field emitted  by an OPA  (Fig.5). Both signal and idler are very noisy and therefore, in the 
case of large photon number, if the object is weak and we detect only the signal field, 
the signal-to-noise ratio for the object is low. 
But, because of the spatial entanglement, the fluctuations in the 
intensity difference between signal and idler are small. Hence if we detect the intensity difference, 
the signal-to-noise ratio for the object becomes much better. This scheme is conceptually related 
to  well known \'' ghost image " experiments \cite{[13],[14]} in the regime  of detection of single photon 
pairs.\\
Next, let us pass to the case of a weak phase object in which one can exploit, instead, the property 
of spatially multimode squeezing. The configuration is the standard one of a  Mach-Zender 
interferometer in which, as it is well known, one can detect a small phase shift with a sensitivity 
beyond the standard quantum limit by injecting  a squeezed beam in the port through  which usually 
normal vacuum enters. If we have a weak phase image (Fig.6) we can obtain the same result by 
injecting a spatially multimode squeezed light \cite{[15]}.

\subsection{Quantum imaging with entangled photon pairs }
\label{qim}

This approach has been formulated by Abouraddy, Saleh, Sergienko,
and Teich for the regime in which single photon pairs are
detected. They have shown that using entangled photons in an
imaging system offers possibilities that cannot be
attained when entanglement is absent \cite{[16]}.\\
In a classical imaging configuration one has a source, an
illumination system, an object, an imaging system and a detector.
The field at the detection plane is related to that at the object
plane by a linear integral transformation with a kernel $h$. Let
us focus on the interesting case illustrated in Fig. 7,
in which the imaging is performed
using photon pairs, in which case the topography of the system
allows for two branches with different kernels to be
simultaneously illuminated, a system that was heuristically
considered by Belinskii and Klyshko \cite{[16a]}. The two-photon
wave function is denoted $\psi(\vec{x}, \vec{x}')$ ($\vec{x}$ and $\vec{x}'$ are
source position vectors in the transverse plane) and the two
kernels are denoted $h_1$ and $h_2$ for the signal and idler beams
1 and 2, respectively. Next, let us introduce two assumptions. The
first is that the two photons are entangled. This is expressed by
the fact that the wave function $\psi$ includes a factor
$\delta(\vec{x}-\vec{x}')$, which expresses the fact that the two photons
are generated at the same point. The second assumption is that the
first photon is detected by a bucket detector, which can reveal
its presence, but not at all its location. On the other hand, the
other detector is capable of scanning the position of the 
photon, and one detects the coincidences  between photon pairs.
The key point is that in this situation the probability
distribution for the position of the second photon depends not only
on $h_2$ but also on $h_1$, and this is true even though the
position of photon 1 is not detected. Thus, suppose that, for example, there
is nothing in the path of photon 2, whereas in the path of photon
1 there is an object which may be transmissive/reflective,
diffusive or scattering (see Fig.7). 
By using a bucket detector for the first
photon, and by scanning the position of the second photon and
detecting coincidences one can obtain an image of the object which
is in the path of the first photon despite the fact that the
position of this photon is not scanned. The essential point is
that, when the two photons are entangled, such distributed quantum
imaging is in general partially coherent, and can possibly be
fully coherent in which case phase information about the object is
preserved , whereas for classically correlated but unentangled
photons the imaging would be incoherent. These authors have also
applied the principle of entangled-photon imaging to quantum
holography \cite{[16]}.

\subsection{Entangled two-photon microscopy }
\label{micro}
In recent years there has been a lively interest in entangled-photon microscopy \cite{[17]}. 

The surge in the development of fluorescence microscopy 
based on two-photon excitation using laser light has been driven by the principal advantages 
of this technique over single-photon excitation: a pair of low-energy photons can deposit 
as much energy as a single ultraviolet photon thereby exciting a fluorescent molecule within a 
sample with greater penetration depth, better resolution, and less risk of damage upon absorption 
along the optical path. However, in order to obtain two-photon absorption with a classical light source such as a laser, 
a very large photon-flux density is necessary to place two photons within a small enough volume and time window so that the 
fluorescent molecule can absorb them. In this latter case, a femtosecond-pulsed high-power laser is used directly 
as the source of light, which can produce undesired photodamage of the specimen.

On the other hand, entangled photons generated by the process of spontaneous parametric downconversion in a 
nonlinear crystal comprise intrinsically paired photons within a small volume and short time window. 
In principle, therefore, far smaller photon-flux densities can be used to effect absorption so that the
risk of photodamage to the specimen is reduced (see Fig. 8). Yet other possible advantages arise
from the direct proportionality of absorption to photon-flux density (rather than the quadratic relation 
that holds for ordinary two-photon absorption) and the fact that the sum of the photon energies of each
entangled pair is a constant and equal to the energy of the downconverted pump photon 
whereas the sum energy is far broader for photons from a femtosecond laser). However, one of the principal 
challenges in implementing this form of microscopy is obtaining an entangled-photon flux that is sufficient to 
excite two-photon transitions, which have a limited cross section \cite{[17a]}.

\subsection{Image amplification by parametric down conversion}
\label{ampli}
Let us come back to the configuration of Fig. 1b in the case of a large number of photon pairs. Let 
us assume that now, instead of a plane wave at frequency close to $\omega_s$, we inject a coherent 
monochromatic  image (Fig.9) of frequency  $\omega_s$. Parametric image amplification has been 
extensively studied from a classical viewpoint (see e.g.\cite{[12]}). A basic point in Fig.9 is that, if 
the image is injected off axis, one obtains in the output a signal image, that represents an amplified 
version of the input image, and also a symmetrical idler image. An interesting situation arises if 
one has, in addition to the amplifier, a pair of lenses located at focal distances with respect to the 
object plane, to the amplifier and to the image plane (Fig.10). As it was shown by our group \cite{[9],[11],[18]}, 
in the limit of large amplification the two output images can be considered twin of each other 
even from a quantum mechanical viewpoint. As a matter of fact, they do not only display the same 
intensity distribution but also the same local quantum fluctuations. Precisely, let us consider two 
symmetrical pixels in the two images (Fig.11). It turns out that the intensity fluctuations in the two 
pixels are identical, i.e. exactly correlated/synchronized. On the other hand, the phase fluctuations 
are exactly  anticorrelated. Hence the situation with respect to phase and intensity fluctuations is 
similar to what one obtains by breaking a fossil in two parts (Fig. 12). In fact the two parts have 
the same  \'' intensity ", if one establishes an analogy between intensity and thickness, but they 
have opposite phase, since one is clockwise and the other anticlockwise, one is convex and the 
other concave.   So in this way, from one image one obtains twin images in a state of spatial 
entanglement which involves also the quadrature components $X$ and $Y$, as it was already described 
in the case of pure parametric fluorescence without any signal injection in Sec.\ref{nearfar}.\\     
There is, however, a  negative point that concerns the signal-to-noise ratio.  When the input 
image is injected off axis, this mechanism of amplification is phase insensitive  and therefore, as it is 
well known, it adds 3 dB of quantum noise in the output\cite{[19]}. In order to have noiseless 
amplification, i.e. amplification which  preserves the signal-to-noise ratio, one must inject two 
coherent images symmetrically (Fig.13) \cite{[18]}. In this case one has in the input two identical, but 
uncorrelated images and in the output two amplified images in a state of spatial quantum 
entanglement.  One can prove that this symmetrical configuration is  phase sensitive and, in fact, as 
it was shown  in  \cite{[20],[21]} the amplification can become noiseless. A couple of years ago there was a 
landmark experiment by  Kumar and collaborators\cite{[22]}  which demonstrated the noiseless 
amplification of a simple test pattern.

\subsection{Measurement of small displacements }
\label{displace}

This topic has been recently studied from a quantum mechanical standpoint by Fabre and his group 
\cite{[23]}. They  analyze the displacement of a light beam. According to the Rayleigh diffraction limit, 
the position of the beam can be measured with an error on the order of the beam section. However, 
one can use, rather, a split detector which measures the intensities  $i_1$ and $i_2$  from the two halves of
the beam cross section (Fig.14).\\
If one displaces gradually the beam with respect to the detector and plots the intensity difference, 
one obtains a curve like that shown in Fig.14 and the precision in the measurement of the 
displacement is limited only by noise. The standard quantum limit is given \cite{[24]} by the ratio of the 
Rayleigh limit to the square root of the photon number. In this way  one can measure shifts in the  
sub-nanometer range. In the case of the microscopic observation of biological samples, one 
cannot raise the photon number and therefore it is important to have the possibility of going beyond 
the standard quantum limit by reducing the noise in the intensity difference  $i_1-i_2$  below the 
shot noise level. To achieve this, a possibility is to use a beam in a state of spatial entanglement 
between its upper and its lower part, because in this way the number of photons in the two parts is 
basically the same and the fluctuations in the intensity difference are very small.
Another possibility, 
which was proposed in \cite{[23]}, is to synthesize a special two-mode state which arises from the 
superposition of a Gaussian mode in a squeezed  vacuum  state and a spatially antisymmetrical mode which 
lies in a coherent state. The antisymmetrical mode is obtained from a Gaussian mode by flipping 
upside down one of the  two halves. One shows that in this way one can measure displacements 
beyond the standard quantum limit. Experiments with this configuration are presently conducted in 
the laboratory of Bachor at the University of Canberra in collaboration with the group of Fabre \cite{[25]}.

\subsection{Image reconstruction}
\label{imaging}
Kolobov and Fabre have recently studied the quantum limits in the process of image reconstruction 
\cite{[26]}. The very well known scheme they analyzed is shown in Fig.15. The object is contained in a 
finite region of size X. The imaging system is composed by two lenses and  an aperture. Because 
of the transverse finiteness of the imaging system, diffraction comes into play, with a consequence 
that some details of the object are blurred. In order to better reconstruct  the object, one can 
proceed as follows. One considers the linear operator H which transforms the object into the image, 
its eigenvalues and its eigenfunctions, which are called prolate spheroidal functions:
\begin{equation}
H f_n (\vec{x}) = \mu_n f_n (\vec{x})
				\label{eigenvalue}
\end{equation}                        
If the imaging were perfect, $H$ would be the identity operator and all eigenvalues would be equal to 
unity. An imperfect imaging introduces a sort of loss  so that  $\mu_n \le 1$.
Now one expands the image $e(\vec{x})$ on the basis of the eigenfunctions:  
\begin{equation}
e(\vec{x})= \sum_n c_n f_n (\vec{x}) 
				\label{imageexp}
\end{equation}                                                                   
and obtains the coefficients $c_n$.  The object $a(\vec{x})$ is reconstructed by the following expression
\begin{equation}
e(\vec{x})= \sum_n \frac{c_n} {\sqrt{\mu_n}} f_n (\vec{x}) 
				\label{}
\end{equation}
where the coefficients $c_n$  have been replaced by $c_n \ \sqrt{\mu_n}$.\\                                    
In principle the reconstruction is perfect, but a problem  is introduced by the circumstance that 
when the index $n$ is increased the eigenvalues $\mu_n$  quickly approach zero. As it is shown in \cite{[26]}, 
this feature makes the reconstruction of fine details very sensitive to noise, so that again quantum 
noise represents the ultimate limitation. The idea proposed in \cite{[26]} is to illuminate the object by 
bright squeezed light instead of coherent light, and to replace the vacuum state  in the part  of the 
object  plane outside the object itself by squeezed vacuum radiation. In this way one can obtain a 
definite improvement of the resolution in the reconstruction, beyond the standard quantum limit.
\subsection{Quantum optical lithography }
\label{qlitho}

This topic was pioneered by Scully and Rathe \cite{[27]} and has become a focus of attention after the 
theoretical paper  by Dowling and collaborators \cite{[28]}. In this work there are two key points (Fig. 
16). First, the entangled photon pairs emitted by an  OPA in a regime of pure parametric down-
conversion are conveyed to a 50/50 beam splitter, and second the lithographic interference pattern 
is obtained by two-photon absorption. The effect is that the wavelength of the interference 
modulation is halved with respect to the standard  one-photon absorption pattern. This is due to the 
fact that, according to a well known  experiment by Hong and Mandel \cite{[29]}, the entangled photons 
emerge from the beam splitter either in pair on the upper side or in pair on the lower side of the 
beam splitter, and there is never a coincidence between one photon on each side. It is just the 
quantum interference between the two possible two-photon paths which produces the halved 
interference pattern. A very recent experiment by Shih and collaborators \cite{[30]}, related to this 
configuration, provides a first-principle demonstration of quantum lithography and exhibits
nicely the halving effect in  the passage from one-photon to two-photon absorption.
A recent article \cite{[31]} considers, instead of the regime of single photon pairs detection, the case in 
which the OPA produces a large number of photons pairs. It shows that in this limit the halving 
effect is well preserved, even if the fringe visibility is reduced by a factor 5. A simple interpretation 
of the halving effect in this case is obtained by observing that the two entangled beams emitted by 
the OPA are transformed by the beam splitter into a pair of squeezed beams (see Sec.\ref{connection}), and 
therefore  one detects the two photon absorption of the interference of two squeezed vacuum fields.
In Fig. 17  we see the squeezing ellipses of the two beams. One of the two ellipses rotates with 
respect to the other  corresponding to the variation of the phase $\chi$  (see Fig.16) which represents 
the phase difference among the two paths. Clearly it is not necessary to perform a rotation  of   $2\pi$.                           
to return to the initial configuration. A rotation of  $\pi$   is enough, and this produces the halving of 
the wavelength. This consideration suggests an alternative procedure to attain an experimental
demonstration of the principle of optical lithography: the interference of two vacuum squeezed beams produced by 
two optical parametric oscillators below threshold which share the same pump beam.\\
As it is well known, a minimum uncertainty squeezed vacuum state has the
form \cite{[2]}
\begin{equation}
\exp {\left[\zeta a^2 -\zeta^* \left(a^\dagger \right)^2\right]} | 0 \rangle 
				\label{squeeze}
\end{equation}
where  $\zeta$   is a c-number and  $| 0 \rangle $ is the vacuum state. One can generalize the
argument above by considering a state of the form
\begin{equation}
\exp {\left[\zeta a^N -\zeta^* \left(a^\dagger\right)^N\right]} | 0 \rangle 
				\label{Nsqueeze}
\end{equation}
which can be obtained, for example, by a material with a $\chi^{N}$
nonlinearity, in which a single pump photon is down-converted into
N entangled photons. One can consider the interference of two beams, one of
which is in state (\ref{Nsqueeze}), while the other lies in the state
\begin{equation}
\exp {\left[\zeta \left(a e^{i\chi}\right)^N -\zeta^* \left(a^\dagger e^{-i\chi}\right)^N\right]} | 0 \rangle 
				\label{Nchisqueeze}
\end{equation}
obtained from (\ref{Nsqueeze}) by a rotation of the angle $\chi$  which corresponds to
the phase difference between the two paths plus, possibly, a constant.
Clearly, a rotation of  $\frac{2\pi} {N}$ is enough to return to the initial
configuration and as a consequence the wavelength of the interference
pattern, detected by N-photon absorption, is reduced by a factor N. Of
course, this argument does not predict the visibility of the fringes.
\subsection{Quantum teleportation of optical images }
\label{qfax}
As a final futuristic point, let us consider the quantum teleportation of optical images \cite{[32]}. 
In this case we consider an OPA  without any input image. Nonetheless in the output we have 
two  twin 
images which are very noisy but still in a state of spatial entanglement, and are sent one to the 
station of Alice  and the other to the station of Bob. The teleportation scheme generalizes the one 
formulated  by Vaidman, Braunstein and Kimble \cite{[33],[34]} for quantum teleportation in terms of 
continuous variables. The input image to be teleported lies in an arbitrary quantum state and using 
a 50/50 beam splitter it is combined with one of the two noisy twin images, so that its quantum 
state is completely corrupted at this stage. However, the system as a whole works as a quantum fax 
machine which is capable of reproducing in the output image not only the average properties of the 
input image, but also the details of its quantum state. This is due to the fact that, thanks to the 
perfect correlation between the two noisy entangled twin images, all the noise introduced at the 
initial state is exactly cancelled in the final stage. This process is called \'' holographic quantum 
teleportation" because it shows striking analogies with standard holography, apart from the fact 
that, of course, it works on a quantum level. 

\section*{Acknowledgments}

We thank  Claude Fabre, Mikhail Kolobov, Eric Lantz, Bahaa Saleh, Alexander Sergienko,
Malvin Teich  for giving us permission of including in 
this paper material provided by them. This work is supported by the European FET  Project 
QUANTIM  (Quantum Imaging). 

\newpage
\centerline{\epsfbox{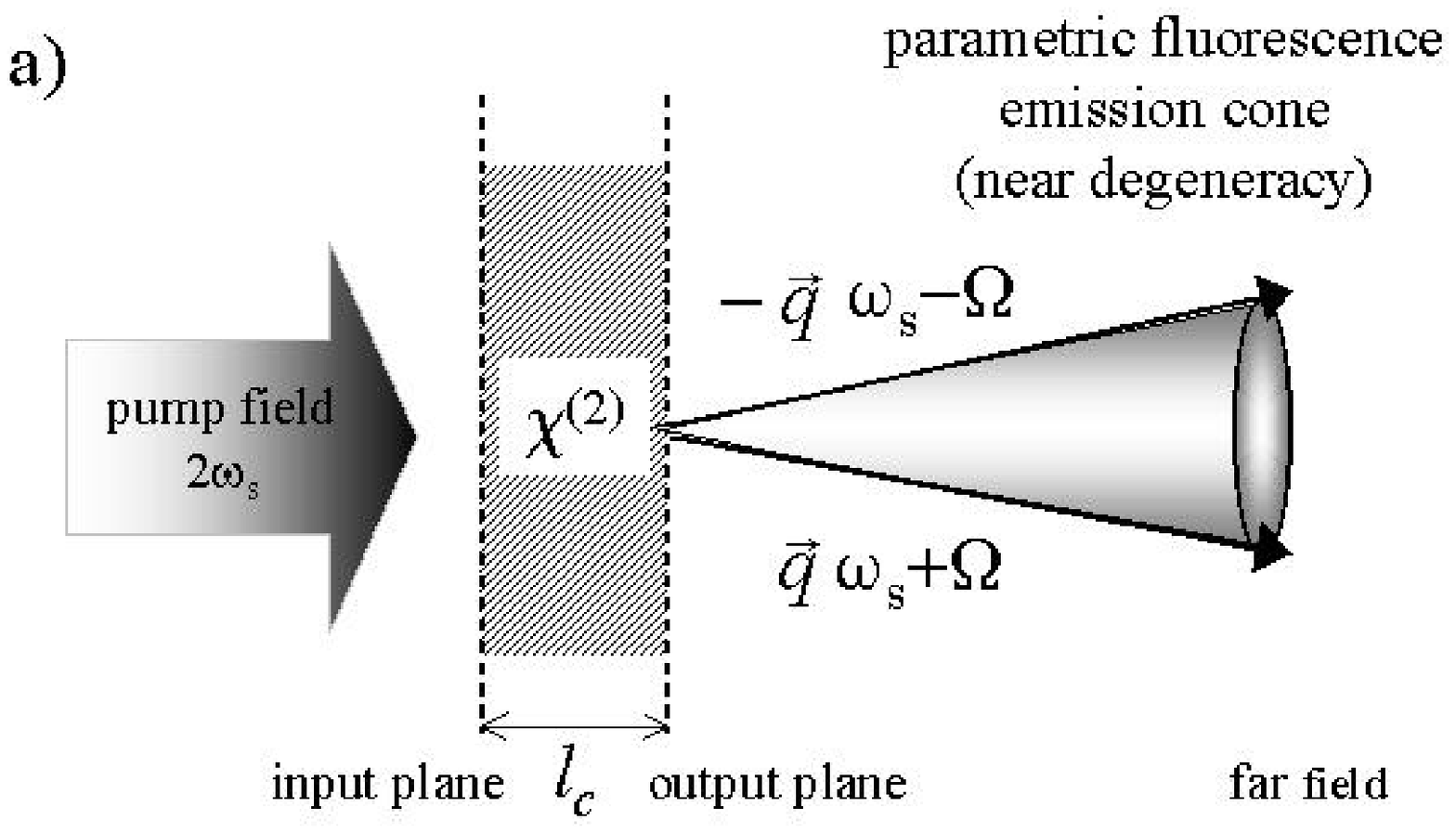}}
\vspace{1.5cm}
\LARGE{Fig.1 (a) Scheme for parametric down-conversion.}

\newpage
\centerline{\epsfbox{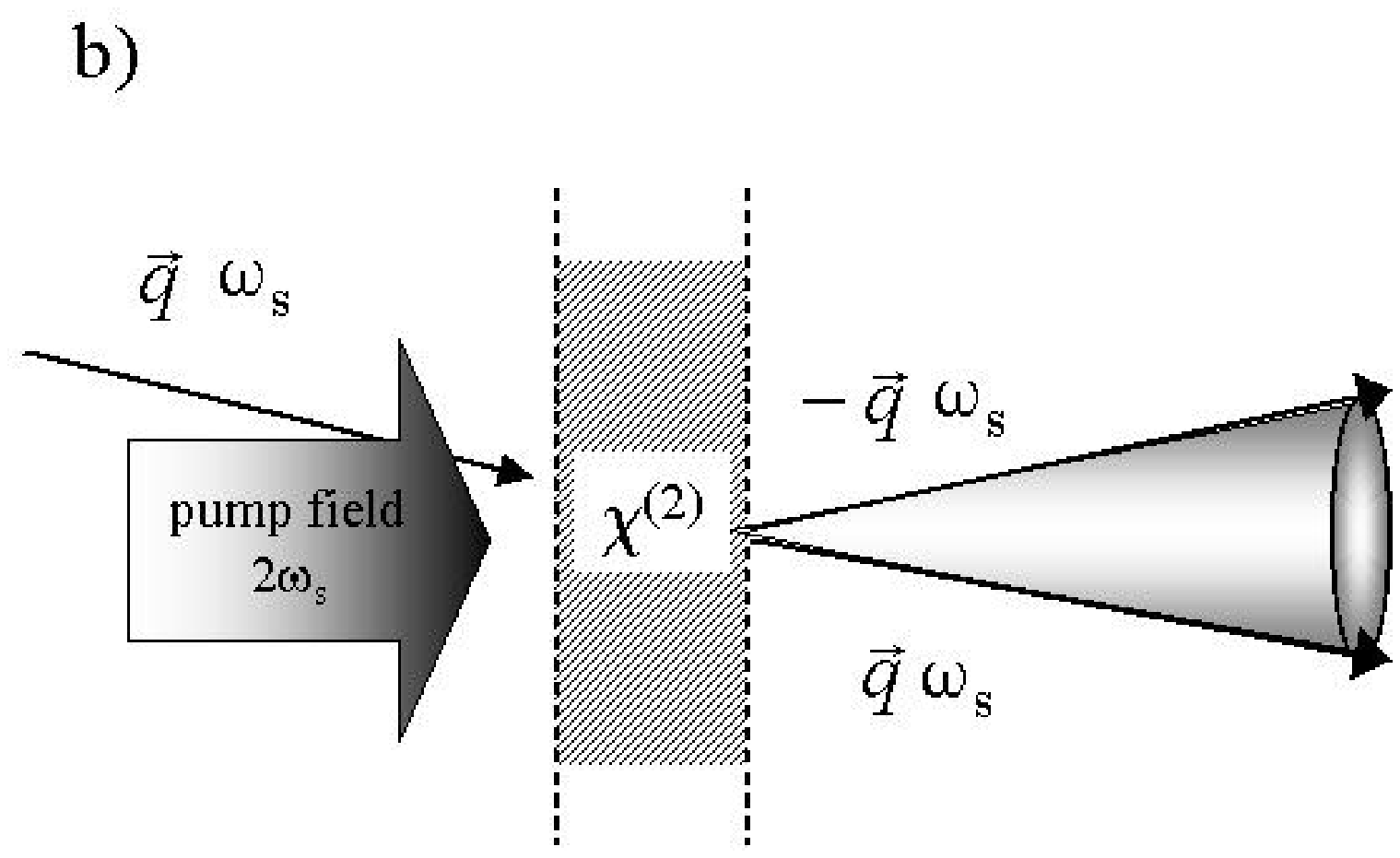}}
\vspace{1.5cm}
\LARGE{Fig 1.(b) Parametric amplification of a plane wave.
$\vec{q}$ is the component of the wave-vector
in the planes orthogonal to the pump propagation direction.} 

\newpage
\centerline{\epsfbox{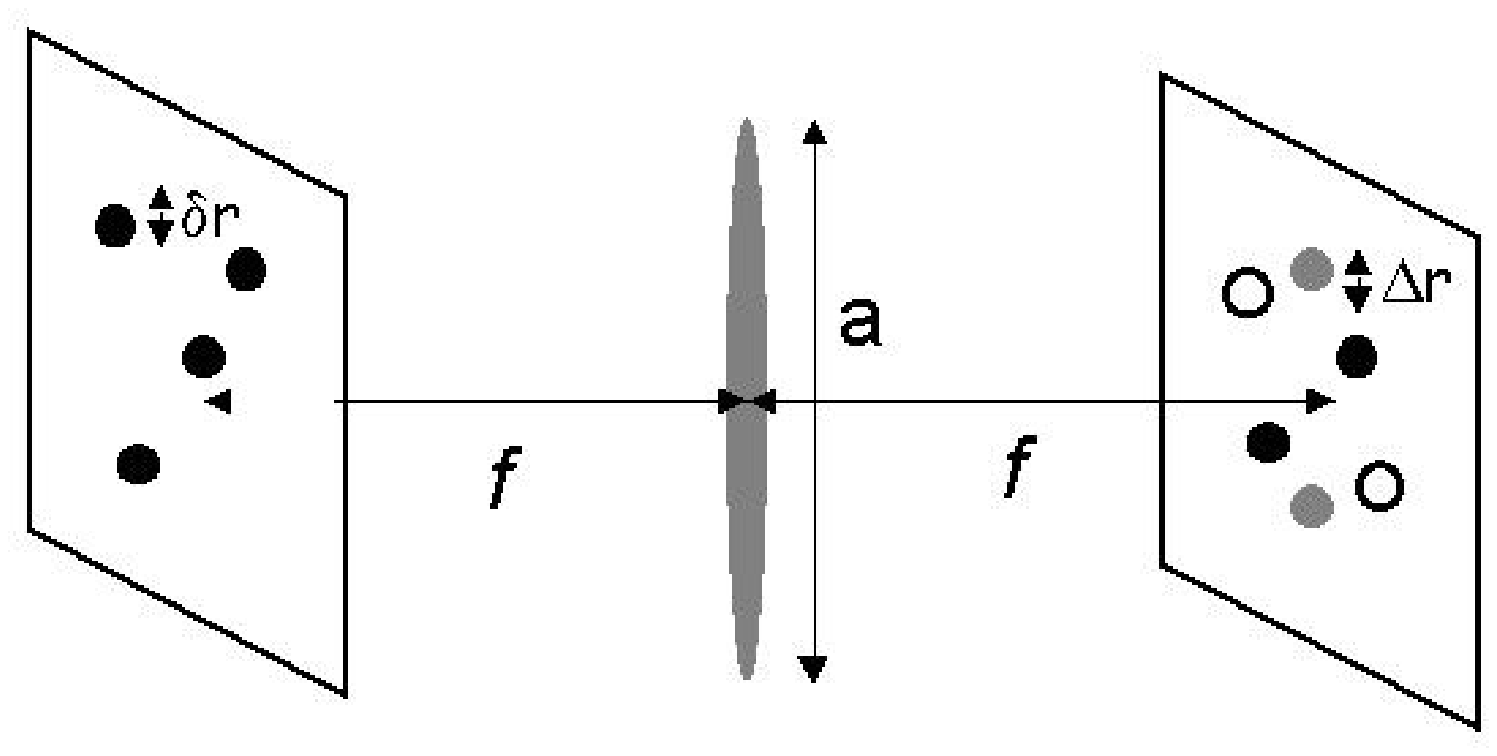}}
\vspace{1.5cm}
\LARGE{Fig.2. Illustration of the near field/far field duality. f is the focal length of the lens.}

\newpage
\centerline{\epsfxsize=10cm \epsfbox{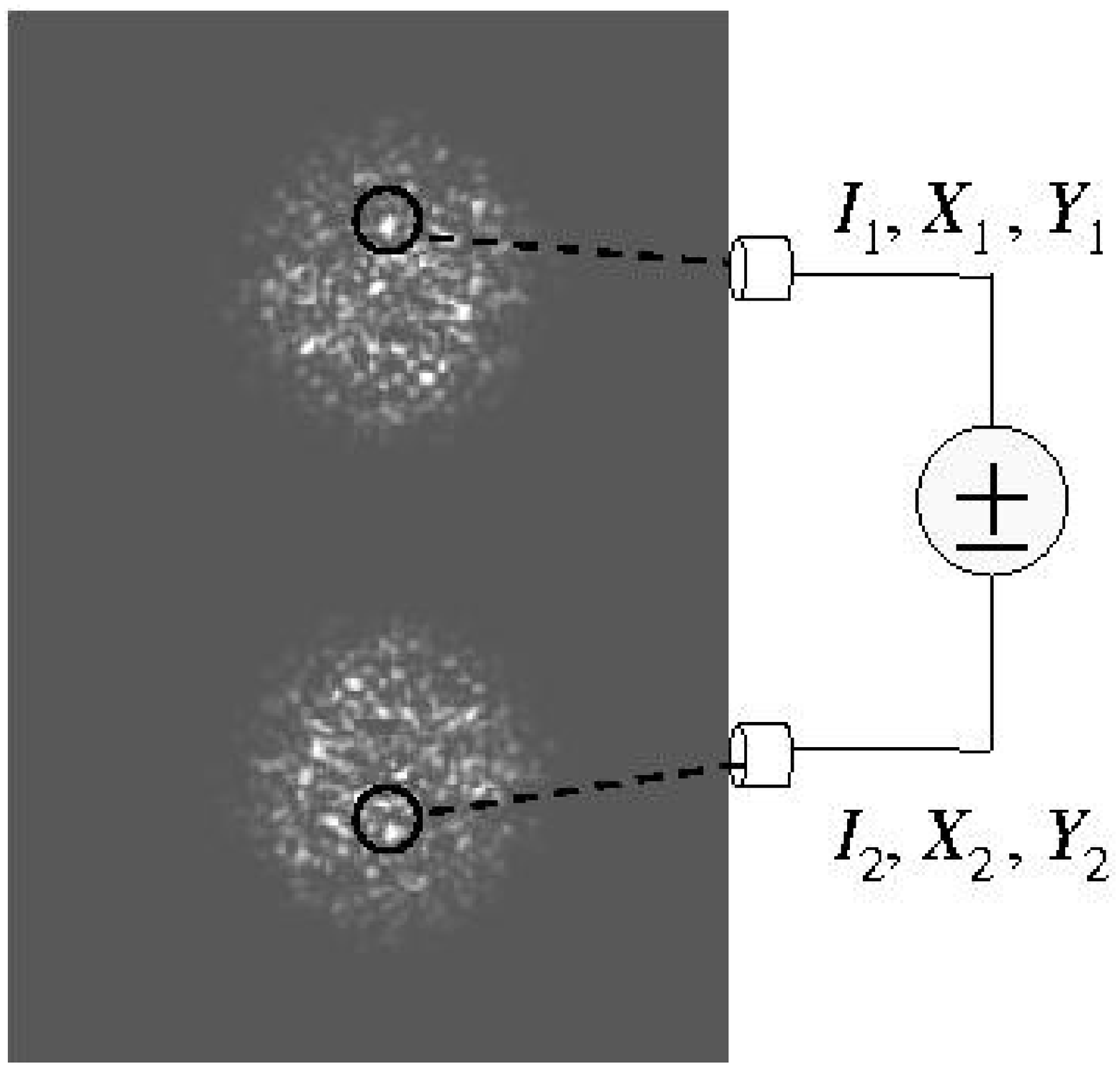}}
\vspace{1.5cm}
\LARGE{Fig.3. Illustration of the concept of spatial entanglement. $I_i$, $X_i$, $Y_i$ $(i=1,2)$ denote the 
intensities and the  quadrature components measured in the two pixels 1 and 2, respectively.}

\newpage
\centerline{\epsfxsize=5cm \epsfbox{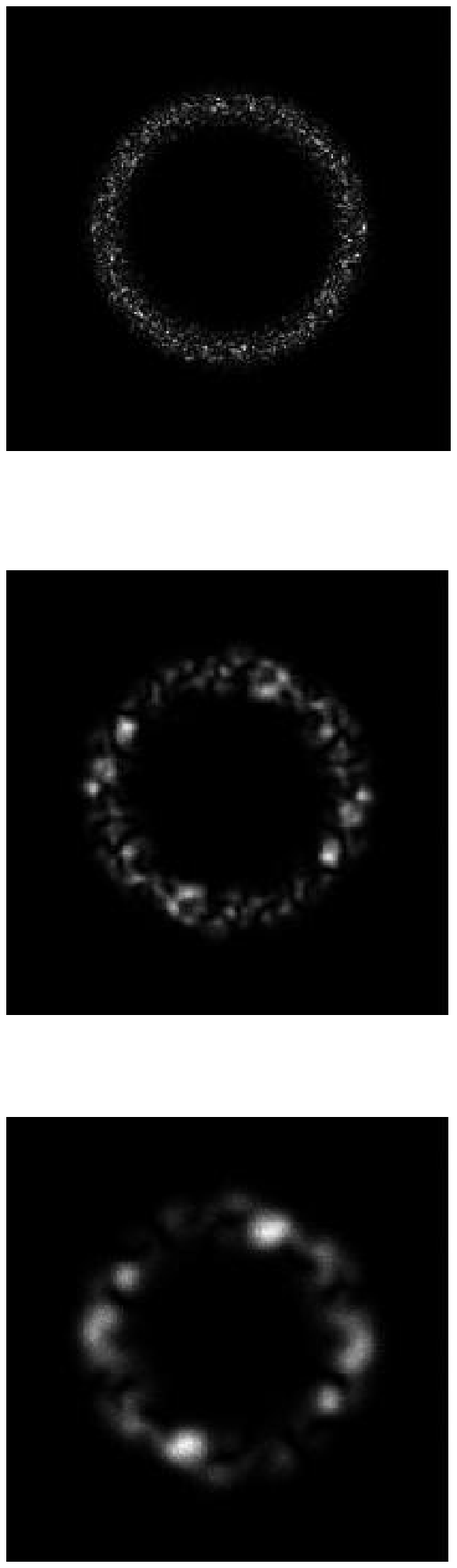}}
\vspace{1.5cm}
\LARGE{Fig.4 (a) Intensity distribution in the far field for a single shot of the pulsed pump field. a) Numerical 
simulations. The waist of the pump beam is 1000 $\mu$m, 300 $\mu$m, 150 $\mu$m 
in the three frames from top to bottom, respectively.}

\newpage
\centerline{\epsfxsize=5cm \epsfbox{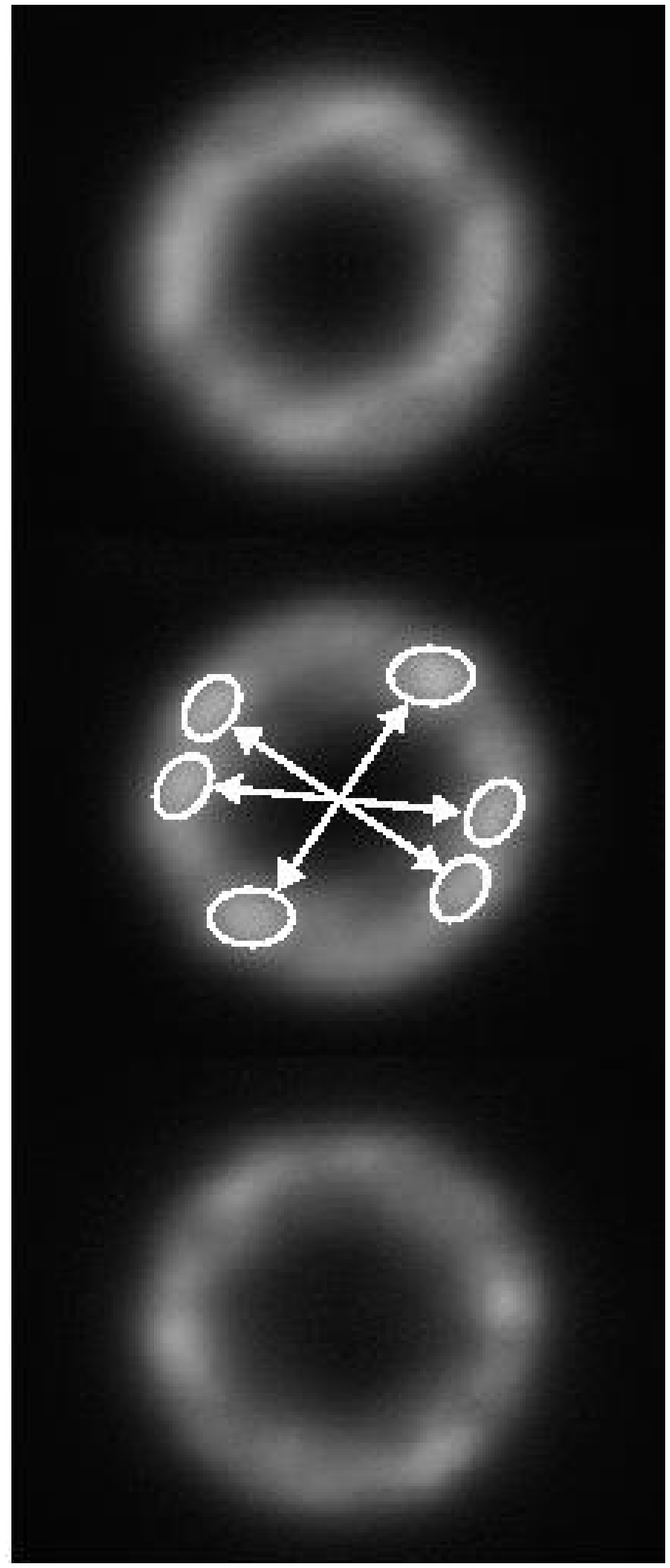}}
\vspace{1.5cm}
\LARGE{Fig.4 (b) Experimental observations by Devaud and 
Lantz at University of Besancon (see [12].}

\newpage
\centerline{\epsfxsize=10cm \epsfbox{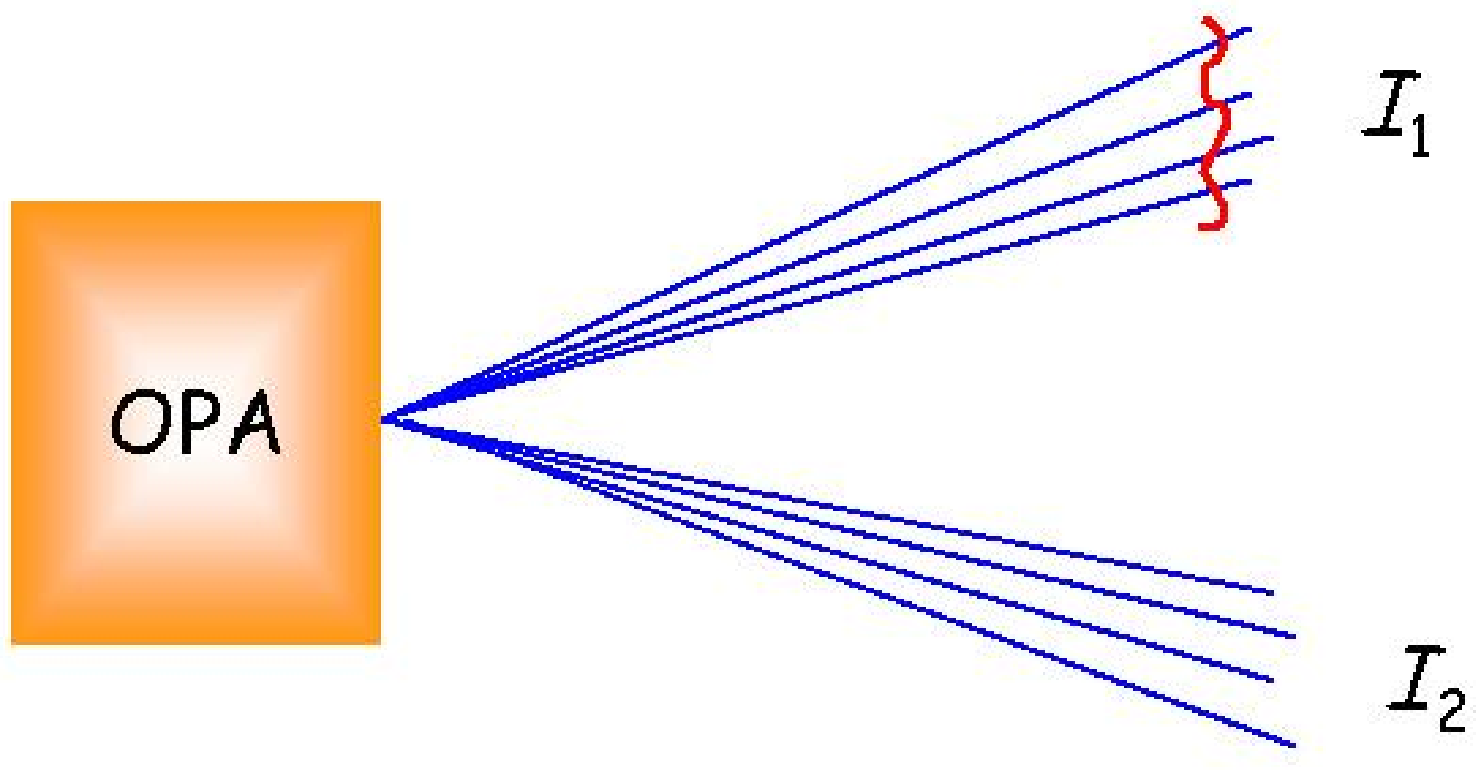}}
\vspace{1.5cm}
\LARGE{Fig.5. Detection of a weak amplitude object by measuring the intensity difference $i_1- i_2$.}

\newpage
\centerline{\epsfbox{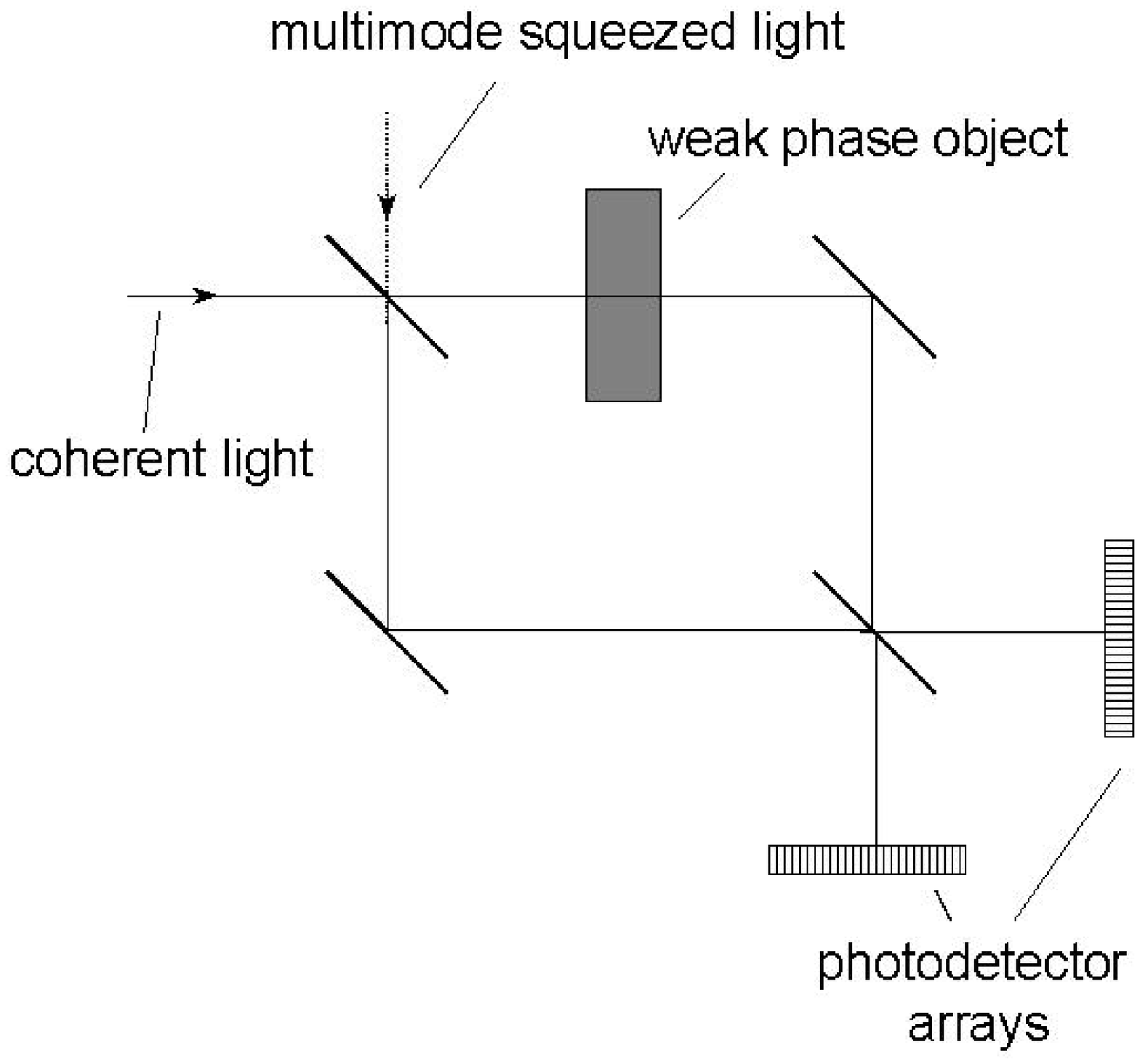}}
\vspace{1.5cm}
\LARGE{Fig.6. Detection of a weak phase object.}

\newpage
\centerline{\epsfxsize=12cm \epsfbox{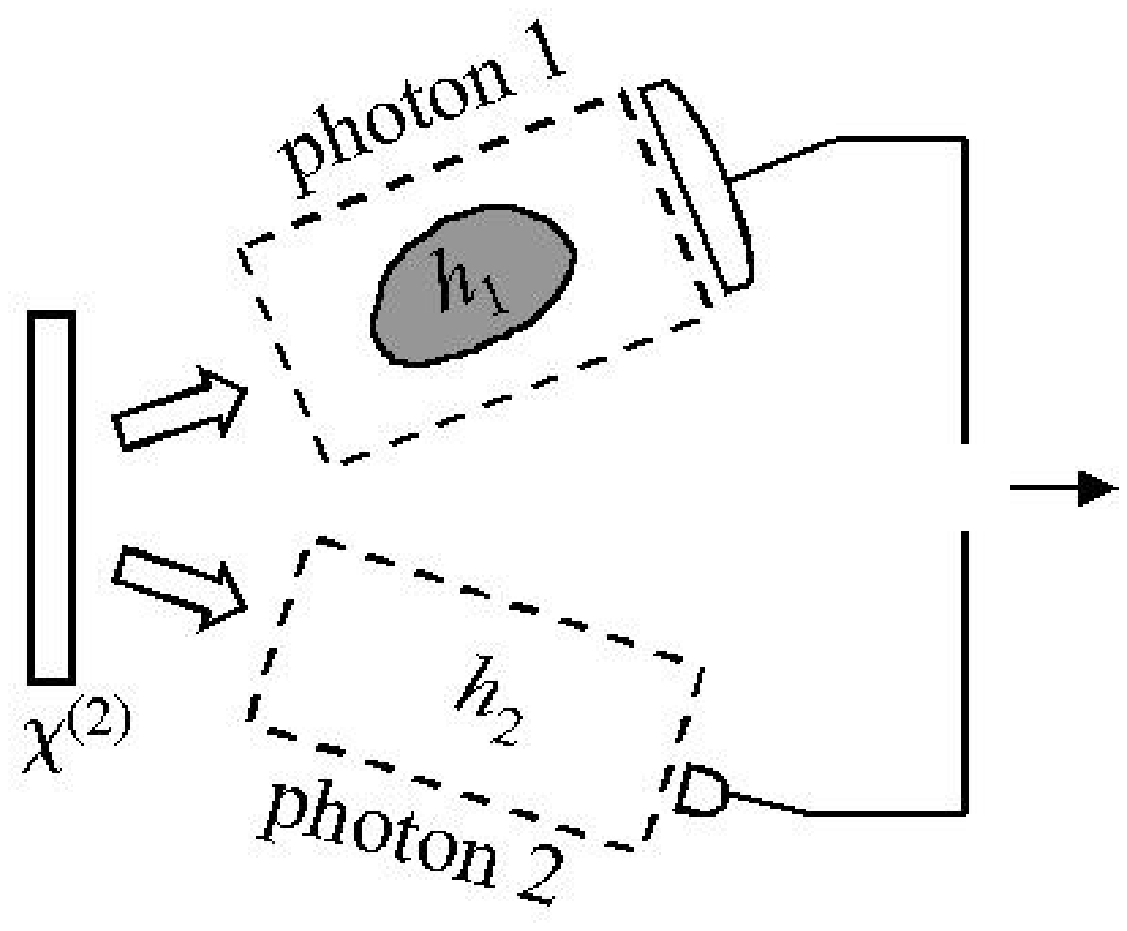}}
\vspace{1.5cm}
\LARGE{Fig.7 . Quantum imaging with entangled photon pairs. 
	Photon 1 is revealed by a bucket detector which does not reveal its transverse position; photon
	2 is observed by a detector that scans its position in coincidence with photon 1 detection.}

\newpage
\centerline{\epsfxsize=12cm \epsfbox{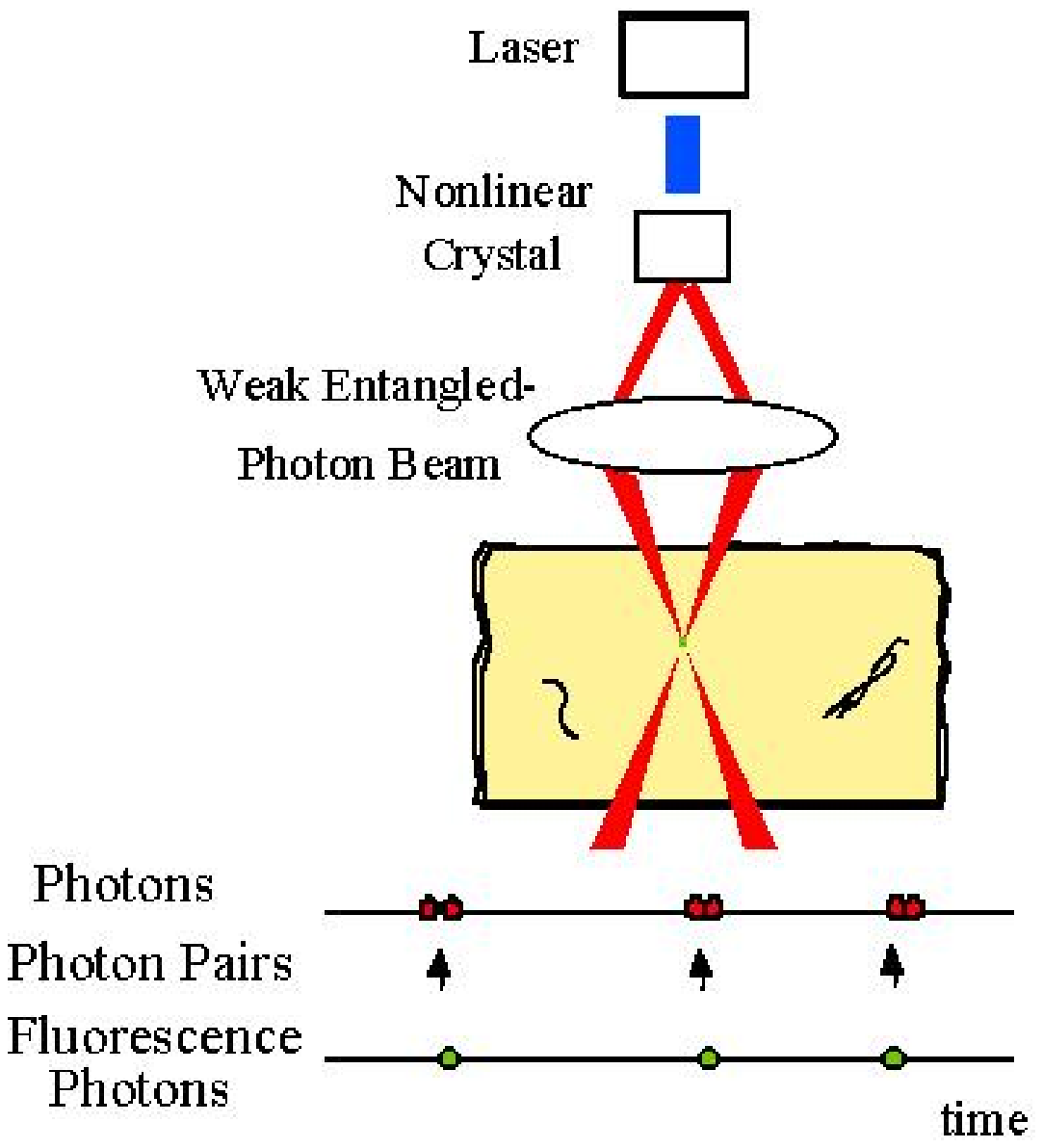}}
\vspace{1.5cm}
\LARGE{Fig.8 Illustration of entangled two photon microscopy.}

\newpage
\centerline{\epsfxsize=12cm \epsfbox{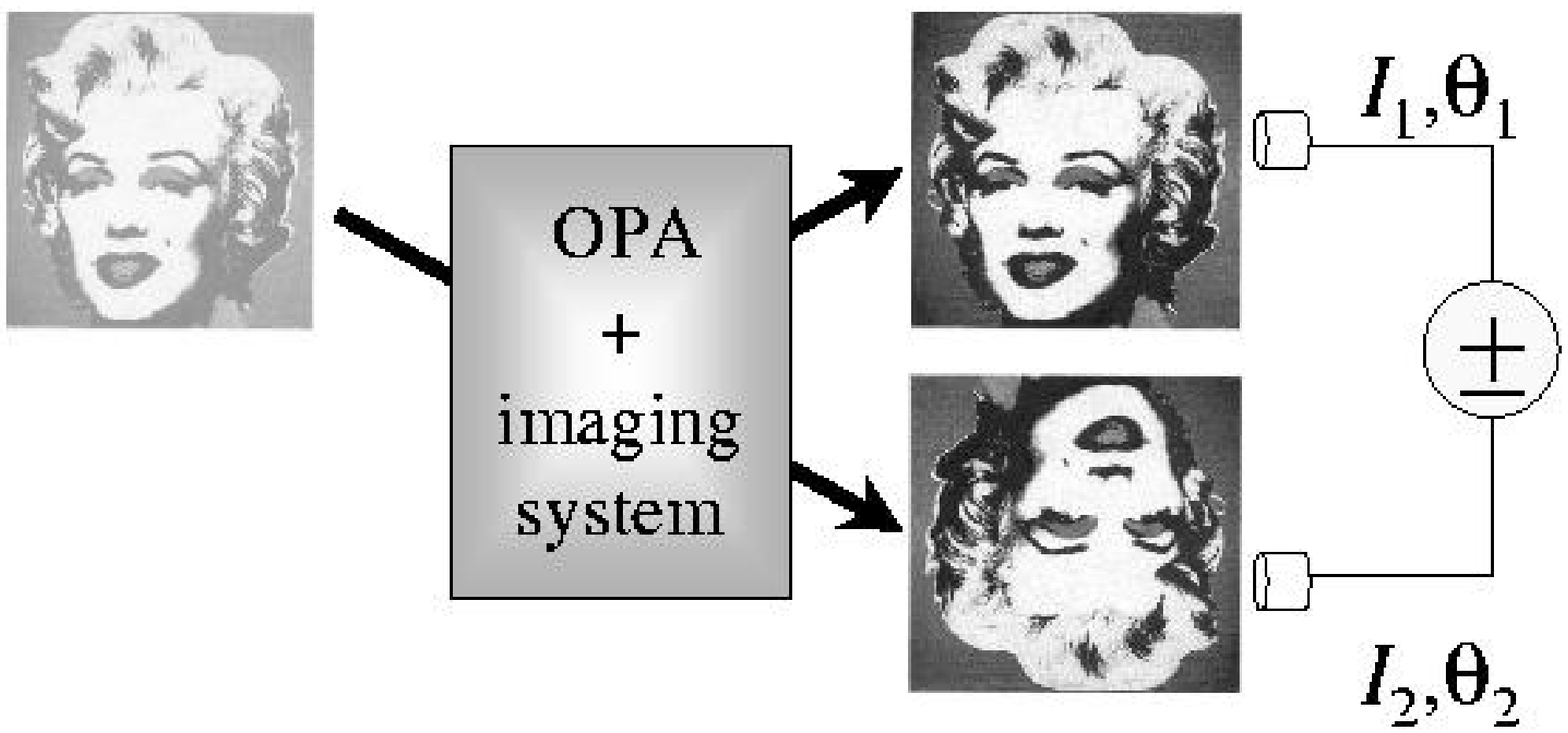}}
\vspace{1.5cm}
\LARGE{Fig.9. Off-axis injection of an image and generation of twin entangled images.}

\newpage
\centerline{\epsfbox{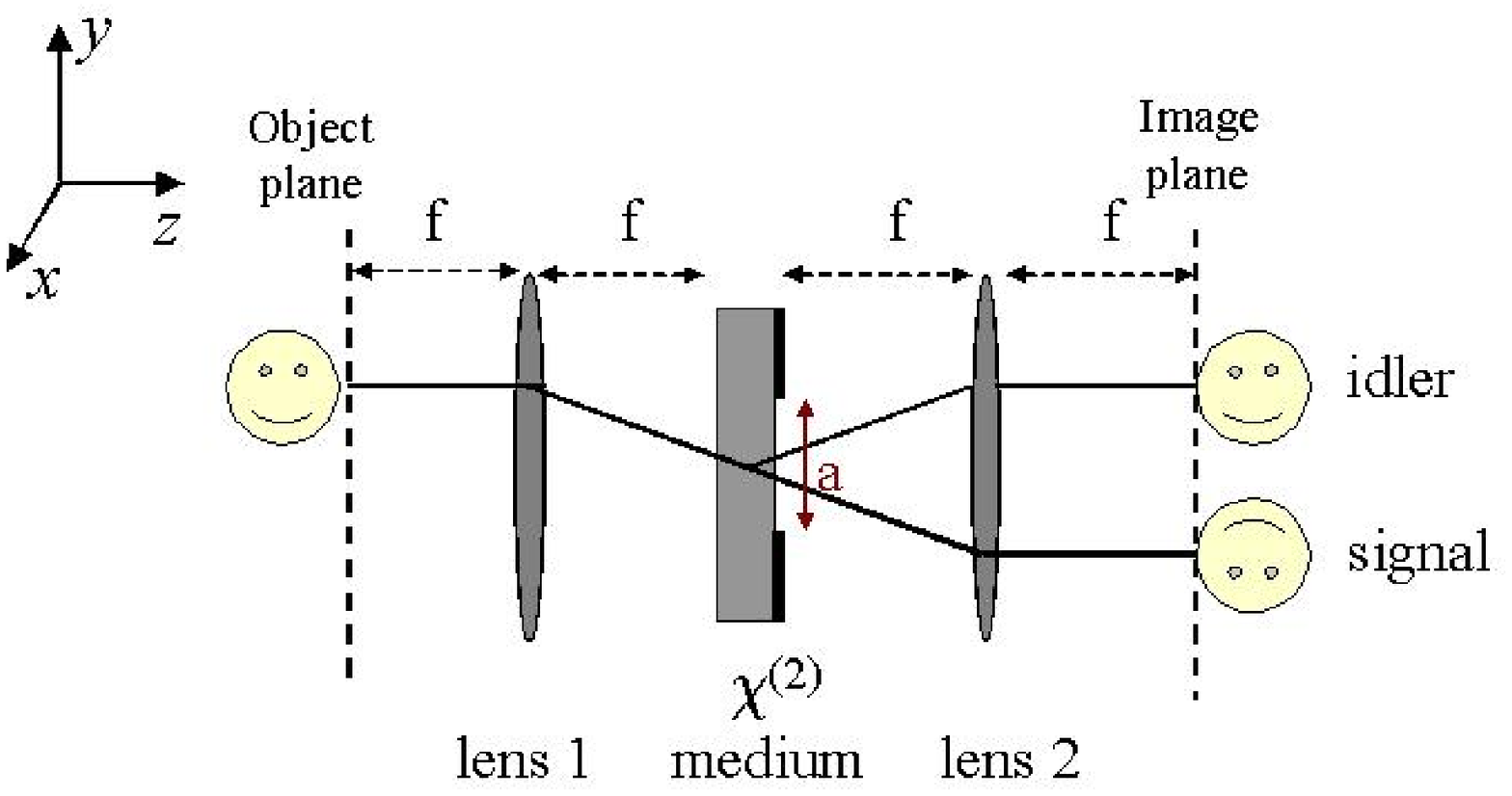}}
\vspace{1.5cm}
\LARGE{Fig.10. Scheme of the parametric optical image amplifier. 
Not shown in the figure is the pump field of frequency $2\omega_s$}

\newpage
\centerline{\epsfxsize=8cm \epsfbox{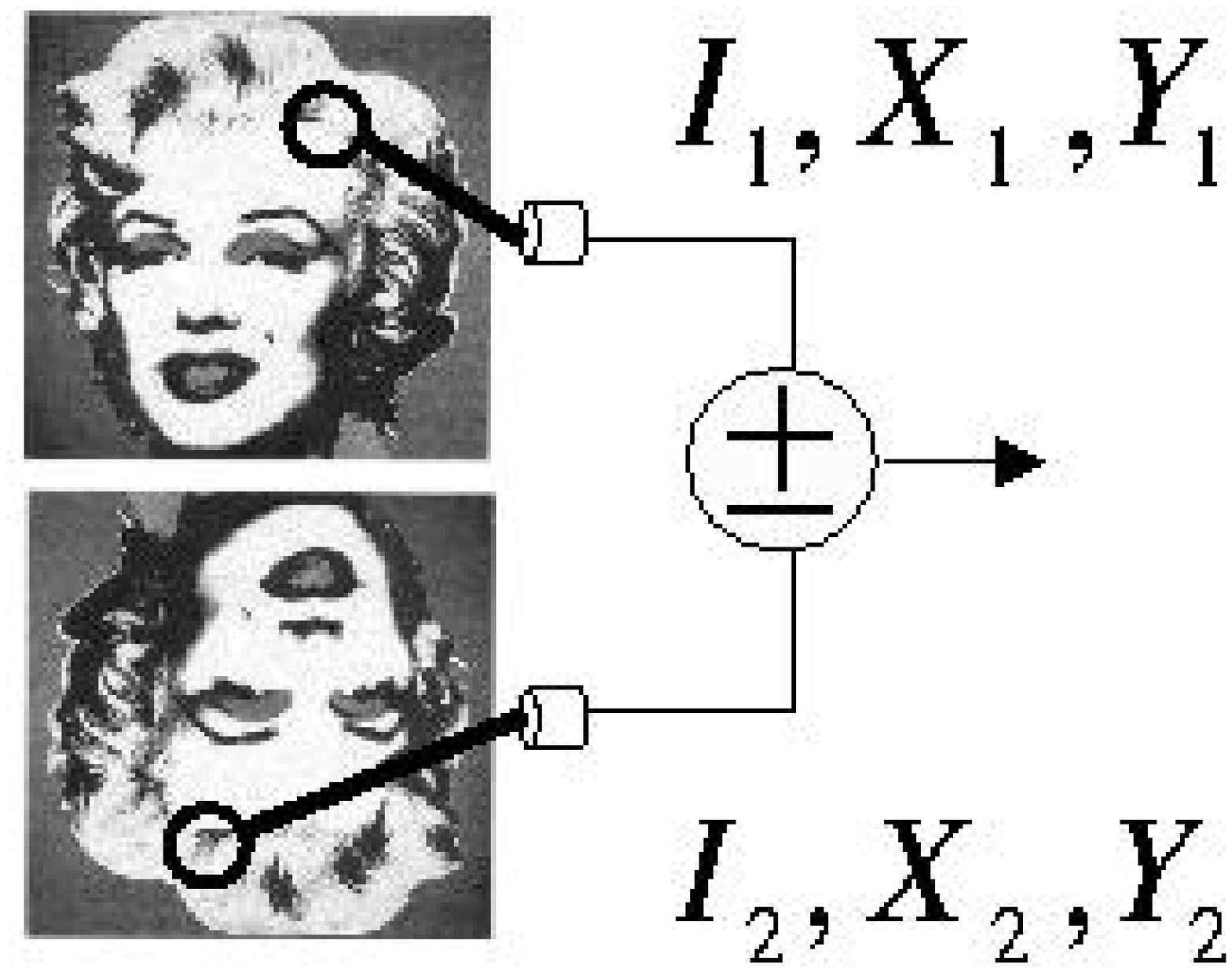}}
\vspace{1.5cm}
\LARGE{Fig.11. The spatial entanglement between the two output images concerns intensity and phase 
fluctuations, and also the fluctuations of quadrature components.}

\newpage
\centerline{\epsfxsize=8cm \epsfbox{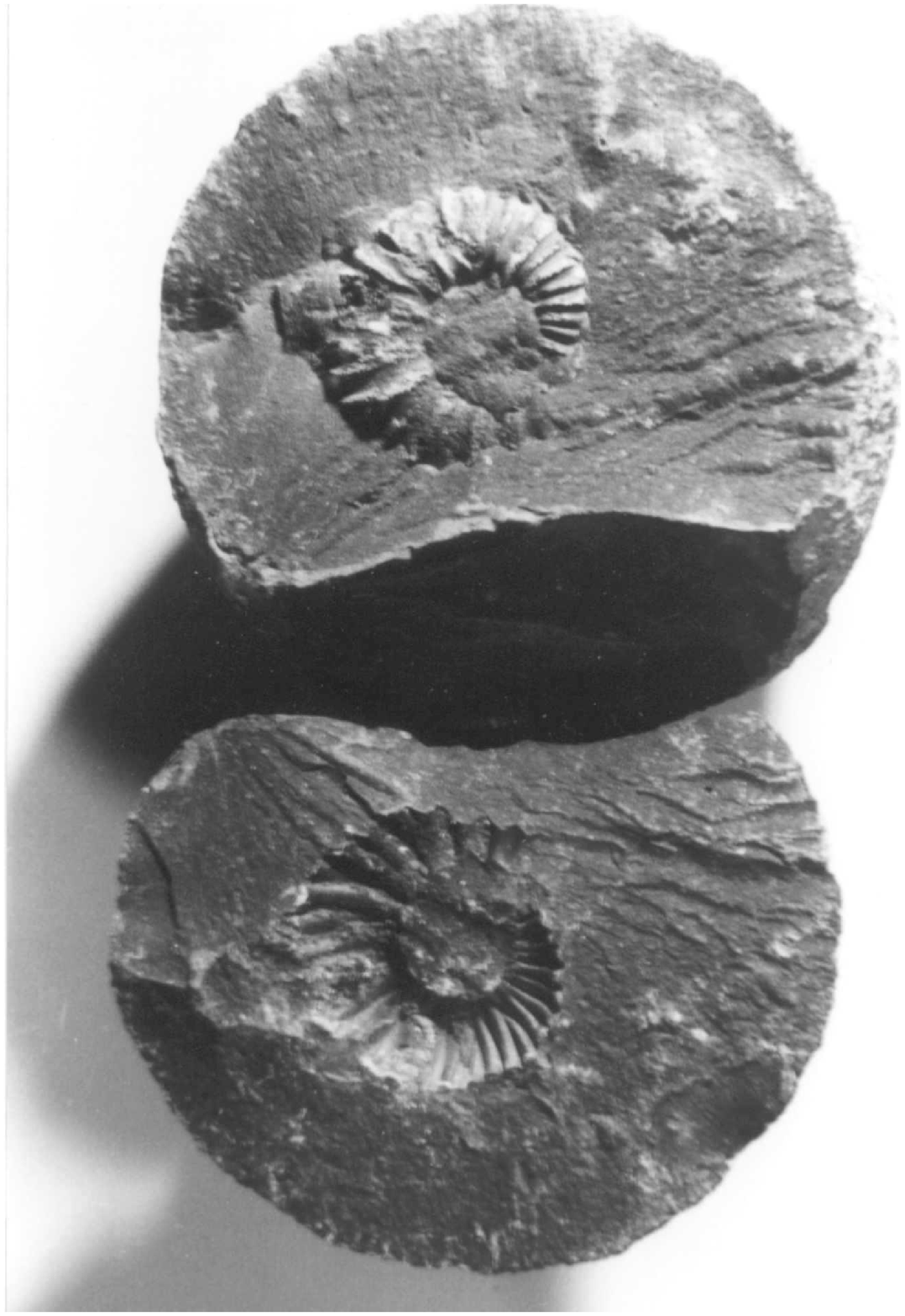}}
\vspace{1.5cm}
\LARGE{Fig.12. Analogy between the twin images and the two parts of a fossil.}

\newpage
\centerline{\epsfbox{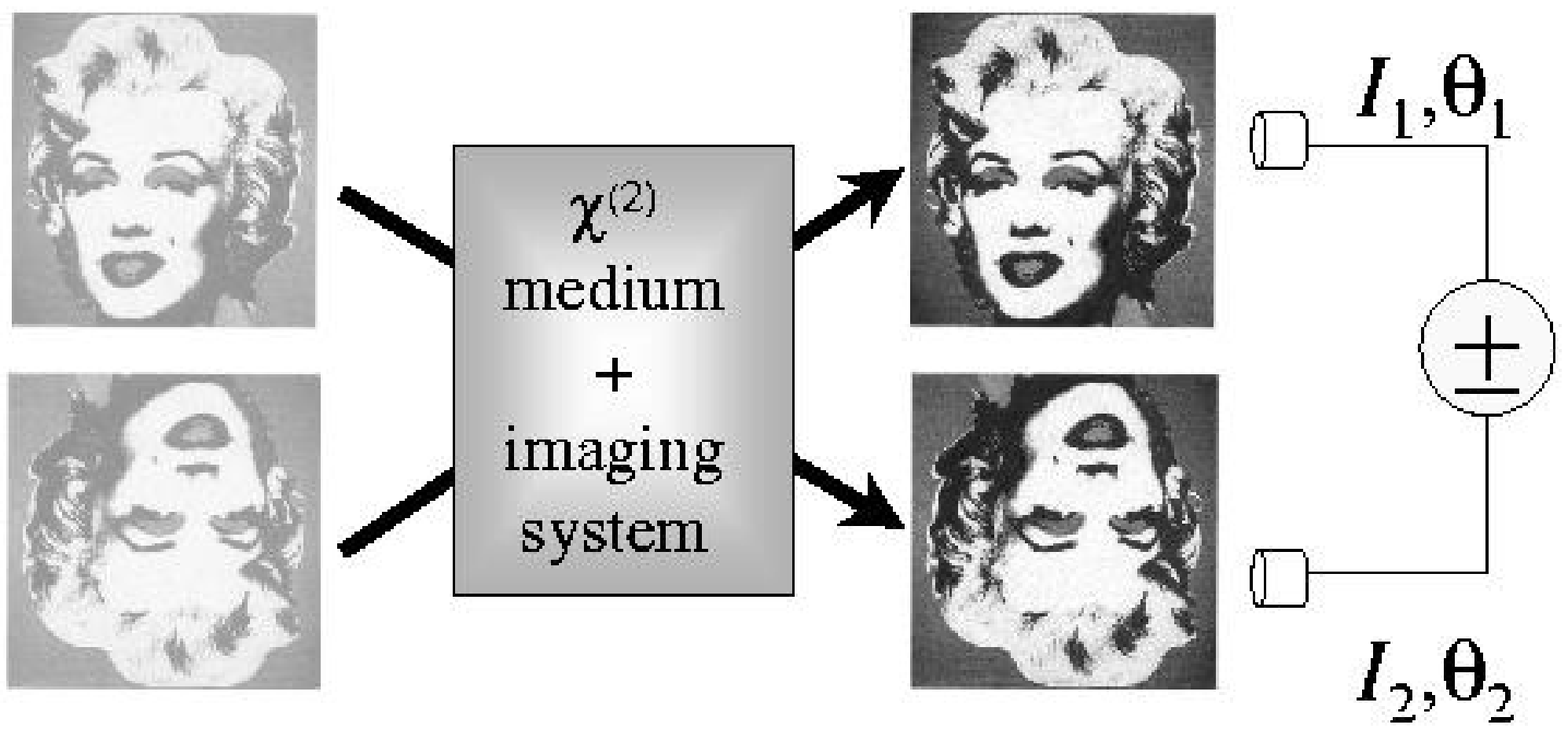}}
\vspace{1.5cm}
\LARGE{Fig.13. Symmetrical injection of an image.}

\newpage
\centerline{\epsfbox{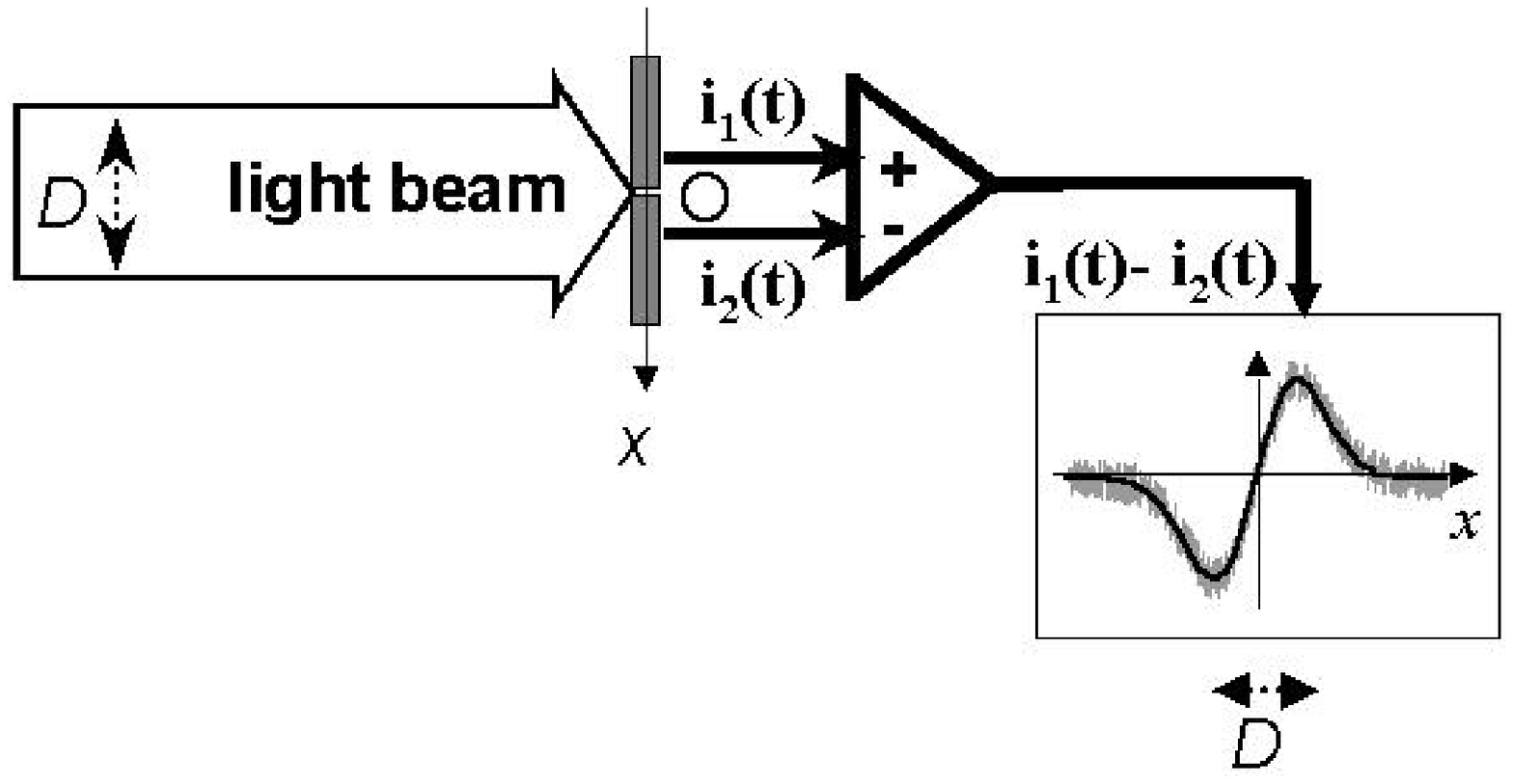}}
\vspace{1.5cm}
\LARGE{Fig.14. Measurement of very small beam displacements in the transverse plane.}

\newpage
\centerline{\epsfbox{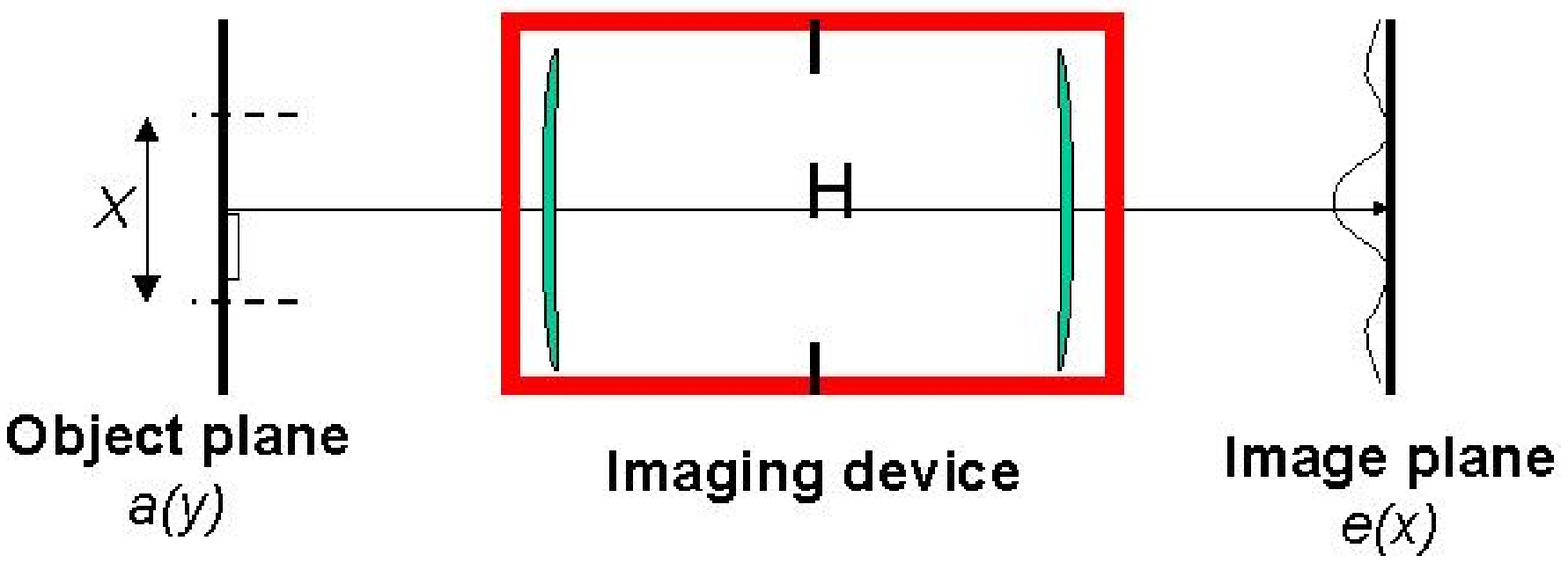}}
\vspace{1.5cm}
\LARGE{Fig.15. Scheme of the imaging system.}

\newpage
\centerline{\epsfbox{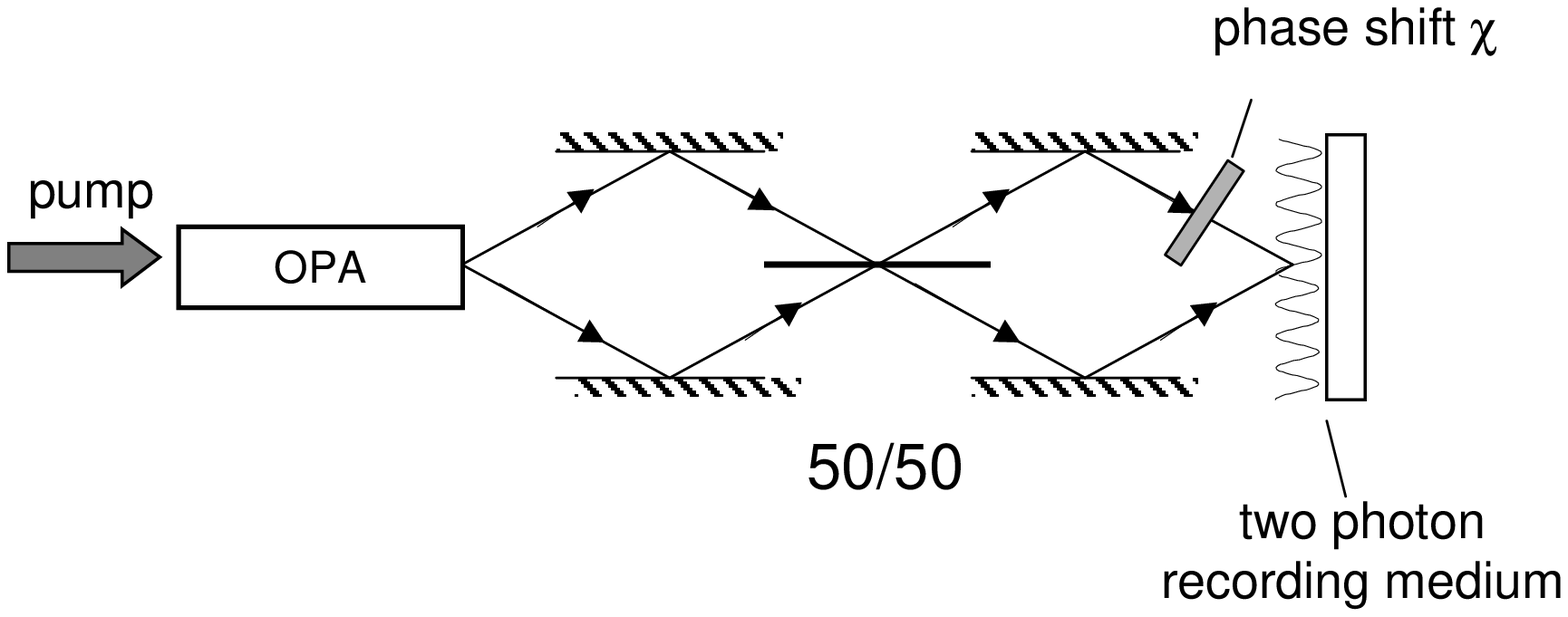}}
\vspace{1.5cm}
\LARGE{Fig.16. Scheme for quantum optical lithography.}

\newpage
\centerline{\epsfbox{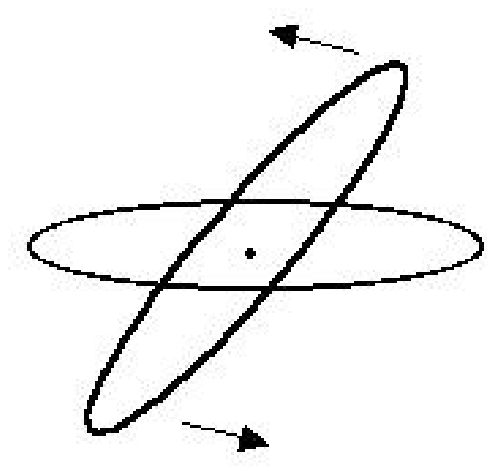}}
\vspace{1.5cm}
\LARGE{Fig.17. Interpretation of quantum lithography as interference between two squeezed vacuum beams.}


\section*{References}
\begin{thebibliography}{100}
\bibitem{[1]} Einstein A., Podolsky B. and Rosen N., Phys.Rev. 47, 777 (1935)
\bibitem{[2]} Walls D.F. and Milburn G., {\em Quantum Optics} (Springer-Verlag, Berlin  1994);
Scully M.O. and Zubairy M.S.,{\em Quantum Optics}
 (Cambridge University Press 1997);Schleich W. {\em Quantum Optics in Phase Space}
(Wiley VCH, 2001) ;
Barnett S.M. and Radmore P.M., {\em Methods in Theoretical Quantum Optics}
(Clarendon Press Oxford 1997)
\bibitem{[3]} Reid M.D.and Drummond P.D.,  Phys. Rev. Lett. 60, 2731 (1988);
              Reid M.D., Phys. Rev. A 40, 913 (1989)
\bibitem{[4]} Ou Z.Y., Pereira S. F.,Kimble  H.J. and Peng K.C.,
        Phys. Rev. Lett.68, 3663 (1992); Ou Z.Y.,
        Pereira S.F. and Kimble H.J., Appl.Phys.B 55, 265 (1992)
\bibitem{[5]}  Kolobov M.I. and Sokolov I.V., Sov.Phys JETP 69, 1097 (1989);
        Phys.Lett.A 140, 101 (1989); Europhys.Lett. 15,271 (1991)
\bibitem{[6]}  Kolobov M.I., Rev. Mod. Phys. 71, 1539 (1999)
\bibitem{[7]}  Lugiato L.A. and Gatti A., Phys.Rev. Lett. 70,
            3868 (1993); Gatti A. and Lugiato L.A., Phys. Rev. A 52,
            1675 (1995); Lugiato L.A., Brambilla M. and Gatti A.,
            Optical Pattern Formation, in {\em Advances in Atomic, Molecular
            and Optical Physics}, vol. 40, 229 (Academic Press, Boston 1999)
\bibitem{[8]}  Lugiato L.A. and Grangier Ph., J. Opt. Soc. Am. B  14, 225 (1997)
\bibitem{[9]}  Gatti A., Brambilla E., Lugiato L.A. and Kolobov M.I.,
            Phys. Rev. Lett. 83, 1763 (1999)
\bibitem{[10]}  Brambilla E., Gatti A., Lugiato L.A. and Kolobov M.I., Eur. Phys. J. D
             {\bf 15}, 127 (2001)
\bibitem{[11]} Navez P., Brambilla E., Gatti A. and Lugiato L.A.,
        Spatial entanglement of twin quantum images, Phys. Rev. A., in press
\bibitem{[12]} Devaud F. and Lantz E., Eur. Phys. J. D 8, 117 (2000);
Lantz E. and Devaud F., Numerical  simulation of spatial fluctuations in parametric
        image amplification, Eur. Phys. J. D, in press (2001).
\bibitem{[13]} Strekalov V., Sergienko A.V., Klyshko D.N., and Shih Y.H.,
        Phys.Rev.Lett. 74,3600 (1995)
\bibitem{[14]} Barbosa G.A.,Phys.Rev.A 54,4473 (1996)
\bibitem{[15]} Kolobov M.I. and Kumar P., Opt. Lett. 18, 849 (1993)
\bibitem{[16]}
        Saleh B.E.A., Abouraddy A.F., Sergienko A.V., and Teich M.C.,
        Phys. Rev. A 62, 043816 (2000);
        Abouraddy A.F., Saleh B.E.A., Sergienko A.V., and Teich M.C.,
        Phys. Rev. Lett. 87, 123602 (2001);
        Abouraddy A.F., Saleh B.E.A., Sergienko A.V., and Teich M.C.,
        Optics Exp. 9, 498 (2001);
        Abouraddy A.F., Saleh B.E.A., Sergienko A.V., and Teich M.C.,
        Entangled-photon Fourier optics, J. Opt. Soc. Am. B 19, in press (2002)
\bibitem{[16a]} Belinskii A.V. and Klyshko D.N., Sov. Phys. JETP
        78, 259 (1994)
\bibitem{[17]} Teich M.C. and Saleh B.E.A., {\v C}eskoslovensk{\'y} {\v c}asopis pro fyziku (Prague) 47, 3 (1997)
\bibitem {[17a]} H.-B. Fei, B. M. Jost, S. Popescu, B. E. A. Saleh, and M. C. Teich, Phys. Rev. Lett. 78, 1679 (1997)
\bibitem{[18]} Gatti A., Brambilla E., Lugiato L.A. and Kolobov M.I.,
J.Opt. B Quantum Semiclass.Opt. 2, 196 (2000)
\bibitem{[19]} Caves C.M., Phys. Rev. D 26, 1817 (1982)
\bibitem{[20]} Kolobov M.I. and Lugiato L.A., Phys Rev. A  52,4930 (1995)
\bibitem{[21]} Sokolov I.V., Kolobov M.I. and Lugiato L.A., Phys. Rev. A 60, 2420 (1998)
\bibitem{[22]} Choi S.-K.,Vasilyev M. and Kumar P., Phys.Rev. Lett. 83, 1938 (1999)
\bibitem{[23]} Fabre C.,Fouet J.B., and Maitre A., Opt. Lett. 25,76 (2000)
\bibitem{[24]} Treps N., Effets quantiques dans les images optiques,
    doctoral thesis at Universite' Paris  VI, 2001
\bibitem{[25]} Fabre C. private communication
\bibitem{[26]} Kolobov M.I. and Fabre C., Phys. Rev.Lett. 85, 3789 (2000)
\bibitem{[27]} Rathe U.V.and Scully M.O., Lett. Math. Phys. 34, 297 (1995)
\bibitem{[28]} Boto A.N., Kok P. Abrams D.S., Braunstein S.L.,
    Williams C.P. and Dowling J.P., Phys.Rev.Lett. 85, 2733 (2000)
\bibitem{[29]} Hong C.V. and Mandel L.,Phys.Rev. A 31, 2409 (1985)
\bibitem{[30]} D'Angelo M., Ceckhova M.V. and Shih Y.H.,
    Two-photon diffraction and quantum lithography, preprint quant-ph/0103035
\bibitem{[31]} Agarwal G.S., Boyd R.W., Nagasako E.M. and Butley S.O.,
        Phys. Rev. Lett. 86, 1389 (2001)
\bibitem{[32]} Sokolov I.V., Kolobov M.I.,Gatti A. and Lugiato L.A.,
    Opt. Comm. 193,175 (2001)
\bibitem{[33]}  Vaidman L., Phys.Rev.A. 49, 1973 (1994)
\bibitem{[34]} Braunstein S.L. and Kimble H.J., Phys.Rev. Lett. 80, 869 (1998)
\end{thebibliography}
\end{document}